\documentclass[11pt]{article}
\newcommand{\nc}{\newcommand}
\nc{\bib}{\bibitem}
\nc{\al}{\alpha}
\nc{\g}{\gamma}
\nc{\G}{\Gamma}
\nc{\D}{\Delta}
\nc{\eps}{\epsilon}
\nc{\la}{\lambda}
\nc{\La}{\Lambda}
\nc{\var}{\varphi}
\nc{\pa}{\partial}
\nc{\nn}{\nonumber \\ }
\nc{\hf}{\frac{1}{2}}         
\nc{\dz}{\frac{dz}{2\pi i}}
\nc{\bin}[2]{\left (\begin{array}{c} {#1}\\ {#2} \end{array}\right )}
\nc{\ben}{\begin{equation}}
\nc{\een}{\end{equation}}
\nc{\bea}{\begin{eqnarray}}
\nc{\eea}{\end{eqnarray}}
\nc{\bra}[1]{\langle {#1}|}
\nc{\ket}[1]{|{#1}\rangle}
\nc{\C}{\mbox{\hspace{1.24mm}\rule{0.2mm}{2.5mm}\hspace{-2.7mm} C}}
\nc{\Nat}{\mbox{\hspace{.04mm}\rule{0.2mm}{2.8mm}\hspace{-1.5mm} N}}
\nc{\NP}[1]{Nucl.\ Phys.\ {\bf #1}}
\nc{\PL}[1]{Phys.\ Lett.\ {\bf #1}}
\nc{\CMP}[1]{Commun.\ Math.\ Phys.\ {\bf #1}}
\nc{\PR}[1]{Phys.\ Rev.\ {\bf #1}}
\nc{\PRL}[1]{Phys.\ Rev.\ Lett.\ {\bf #1}}
\nc{\PTP}[1]{Prog.\ Theor.\ Phys.\ {\bf #1}}
\nc{\PTPS}[1]{Prog.\ Theor.\ Phys.\ Suppl.\ {\bf #1}}
\nc{\MPL}[1]{Mod.\ Phys.\ Lett.\ {\bf #1}}
\nc{\IJMP}[1]{Int.\ Jour.\ Mod.\ Phys.\ {\bf #1}}
\nc{\IM}[1]{Invent.\ Math.\ {\bf #1}}
\nc{\SJNP}[1]{Sov. J. Nucl. Phys.\ {\bf #1}}
\nc{\JHEP}[1]{J.\ High\ Energy Phys.\ {\bf #1}}

\def\vvdots{\mathinner{\mkern1mu\raise1pt\vbox{\kern7pt\hbox{.}}\mkern2mu
 \raise4pt\hbox{.}\mkern2mu\raise7pt\hbox{.}\mkern1mu}}

\begin{document}

\topmargin -5mm
\oddsidemargin 5mm

\begin{titlepage}
\setcounter{page}{0}

\vspace{8mm}
\begin{center}
{\huge A non-reductive $N=4$ superconformal algebra}

\vspace{15mm}
{\large J{\o}rgen Rasmussen}\footnote{rasmussj@cs.uleth.ca} 
\\[.2cm]
{\em Physics Department, University of Lethbridge,
Lethbridge, Alberta, Canada T1K 3M4}

\end{center}

\vspace{10mm}
\centerline{{\bf{Abstract}}}
\vskip.4cm
\noindent
A new $N=4$ superconformal algebra (SCA) is presented.
Its internal affine Lie algebra is based on the seven-dimensional Lie
algebra $su(2)\oplus g$, where $g$ should be identified with a 
four-dimensional non-reductive Lie algebra. 
Thus, it is the first known example of what we choose to call a
non-reductive SCA. It contains a total of 16 generators and is obtained
by a non-trivial In\"on\"u-Wigner contraction of the well-known large
$N=4$ SCA. The recently discovered asymmetric $N=4$ SCA is a 
subalgebra of this new SCA. Finally, the possible affine extensions
of the non-reductive Lie algebra $g$ are classified. The two-form governing
the extension appearing in the SCA differs from the ordinary Cartan-Killing
form.
\end{titlepage}
\newpage
\renewcommand{\thefootnote}{\arabic{footnote}}
\setcounter{footnote}{0}

\section{Introduction}

$N=4$ superconformal algebras (SCAs) in two dimensions have been studied
extensively \cite{Aetal,Sch,STV,AK,Ras2}.
Their internal affine Lie algebras are all Kac-Moody (KM) algebras
based on reductive Lie algebras.
We recall that a reductive Lie algebra $g$ admits a decomposition into a
semi-simple Lie algebra and a direct sum of $u(1)$'s:
\ben
 g=g_1\oplus\dots\oplus g_n\oplus \underbrace{u(1)\oplus\dots\oplus u(1)}_m
\een
$g_i$ is a simple Lie algebra, and $n$ or $m$ may be zero.
To the best of our knowledge, all known SCAs have an internal
KM algebra based on a reductive Lie algebra. In the simple case
of only one supercurrent, $N=1$, the internal symmetry group is
generated by the identity, i.e., $n=m=0$.

The internal affine Lie algebra of the small $N=4$ SCA \cite{Aetal} 
is based on $su(2)$; the large $N=4$ SCA \cite{Sch,STV}
on $su(2)\oplus su(2)\oplus u(1)$; the middle $N=4$ SCA 
\cite{AK} on $su(2)\oplus u(1)\oplus u(1)\oplus u(1) \oplus u(1)$;
while the asymmetric $N=4$ SCA \cite{Ras2} is based on 
$su(2)\oplus u(1)\oplus u(1)$.
The total number of generators in the four examples listed above
are 8, 16, 16 and 12, respectively.
Besides the Virasoro generator, the four supercurrents, and the
affine Lie algebra generators, the remaining generators are all 
spin-1/2 fields.

It is possible to extend the asymmetric $N=4$ SCA by considering one of
the $u(1)$ generators as the derivative of a scalar.
It is this slightly bigger SCA that is provided in \cite{Ras2}.

Here we shall consider a non-trivial In\"on\"u-Wigner (IW) contraction
of the large $N=4$ SCA. The resulting algebra is of course closed
under (anti-)commutators, whereas the Jacobi identities are not necessarily
satisfied. However, we have checked explicitly that they are. They read
\ben
 0=(-1)^{ac}[A_r,[B_s,C_t{\}}{\}}+(-1)^{ba}[B_s,[C_t,A_r{\}}{\}}
   +(-1)^{cb}[C_t,[A_r,B_s{\}}{\}}
\label{Jacobi}
\een
where $a$ is the parity of the field $A$ etc. $[\cdot,\cdot{\}}$ denotes an
anti-commutator when both generators are fermionic, i.e., of odd parity.
It is otherwise a commutator.

The internal symmetry group of the resulting $N=4$ SCA turns out to be
generated by an affine Lie algebra based on the seven-dimensional
Lie algebra $su(2)\oplus g$, where $g$ is a four-dimensional 
{\em non-reductive} Lie algebra to be discussed below.
This novel kind of $N=4$ SCA has 16 generators: the Virasoro
generator, $N=4$ supercurrents, the seven-dimensional affine
Lie algebra, and four spin-1/2 fermions.

Affine extensions of non-reductive Lie algebras are in general
not unique. Thus, the affine extension of $g$ that
appears in our $N=4$ SCA is dictated by the complex structure of the full 
SCA. We classify the possible affine extensions of $g$, and find
that the extension relevant to our construction differs from the more
conventional one governed by the ordinary Cartan-Killing form.

The guiding principle while considering the particular and somewhat
peculiar IW contraction alluded to above, was to
look for an extension of the recently discovered asymmetric $N=4$ SCA
\cite{Ras2}. Indeed, the latter appears as a subalgebra of the likewise
asymmetric new and non-reductive
$N=4$ SCA, rendering its unfamiliar number of 12 generators less mysterious.

In the string theoretical context of the $AdS$/CFT correspondence,
some explicit constructions of superconformal algebras have been obtained.
The Virasoro algebra was considered in \cite{GKS}, $N=1,\ 2$ and $4$
SCAs were discussed in \cite{Ito},
while a general approach to constructing SCAs on the
boundary of $AdS_3$ was outlined in \cite{Ras2}. Those works rely on free
field realizations of affine current (super-)algebras on the world sheet
of the string theory. It would be interesting
to see how the new $N=4$ SCA presented here fits into that framework.
Free field realizations of generic affine
current superalgebras first appeared in \cite{Ras1}.

\section{Large $N=4$ superconformal algebra}

Following \cite{STV}, the large $N=4$ SCA is generated by
$L,G^a,A^{\pm i},U$ and $Q^a$ with $a=1,2,3,4$ and $i=1,2,3$.
The conformal weights $\D$ for ${\{}G,A,U,Q{\}}$ are ${\{}3/2,1,1,1/2{\}}$.
$A$ and $U$ generate an affine $su(2)\oplus su(2)\oplus u(1)$ Lie algebra.
The large $N=4$ SCA is a two-parameter (doubly extended)
algebra in terms of $c$ and $\g$ or equivalently $k^+$ and $k^-$:
\bea
 k^+=\frac{c}{6\g}\ ,&&\ \ \ \ \ k^-=\frac{c}{6(1-\g)}\nn
 c=\frac{6k^+k^-}{k^++k^-}\ ,&&\ \ \ \ \ \g=\frac{k^-}{k^++k^-}
\label{kk}
\eea
For convenience of notation, one introduces $4\times4$-matrices 
$\al$ satisfying
\ben
 [\al^{\pm i},\al^{\pm j}]=-\sum_{k=1}^3\epsilon^{ijk}\al^{\pm k}\ ,\ \ \ 
 [\al^{+i},\al^{-j}]=0\ ,\ \ \ \{\al^{\pm i},\al^{\pm j}\}=-\hf\delta^{ij}
\een
They can be represented by
\ben
 \al^{\pm i}_{ab}=\pm\hf\left(\delta^i_a\delta^4_b-\delta^i_b\delta^4_a
   \right)+\hf\epsilon^{iab}
\een
The large $N=4$ SCA may now be written 
\bea
 {[}L_n,L_m{]}&=&(n-m)L_{n+m}+\frac{c}{12}(n^3-n)\delta_{n+m,0}\nn
 {[}L_n,\Phi_r{]}&=&((\Delta(\Phi)-1)n-r)\Phi_{n+r}\nn
 \{G^a_r,G^b_s\}&=&2\delta^{ab}L_{r+s}-4(r-s)\sum_{i=1}^3
   \left(\g\al_{ab}^{+i}A_{r+s}^{+i}+(1-\g)\al_{ab}^{-i}A_{r+s}^{-i}\right)\nn
 &&+\frac{c}{3}(r^2-1/4)\delta^{ab}
    \delta_{r+s,0}\nn
 {[}A^{+i}_n,G^a_r{]}&=&\sum_{b=1}^4\al_{ab}^{+i}\left(G^b_{n+r}-2(1-\g)
    nQ^b_{n+r}\right)\nn
 {[}A^{-i}_n,G^a_r{]}&=&\sum_{b=1}^4\al_{ab}^{-i}\left(G^b_{n+r}+2\g
    nQ^b_{n+r}\right)\nn
 {[}A^{\pm i}_n,A^{\pm j}_m{]}&=&\sum_{k=1}^3\epsilon^{ijk}A^{\pm k}_{n+m}
   -\frac{k^\pm}{2}n\delta^{ij}\delta_{n+m,0}\nn
 {[}A^{\pm i}_n,Q^a_r{]}&=&\sum_{b=1}^4\al_{ab}^{\pm i}Q^b_{n+r}\nn
 {[}U_n,G^a_r{]}&=&nQ^a_{n+r}\nn
 {[}U_n,U_m{]}&=&-\frac{c}{12\g(1-\g)}n\delta_{n+m,0}\nn
 \{Q^a_r,G^b_s\}&=&2\sum_{i=1}^3\left(\al_{ab}^{+i}A_{r+s}^{+i}-
   \al_{ab}^{-i}A_{r+s}^{-i}\right)+\delta^{ab}U_{r+s}\nn
 \{Q^a_r,Q^b_s\}&=&-\frac{c}{12\g(1-\g)}\delta^{ab}\delta_{r+s,0}\nn
 0&=&{[}A^{+i}_n,A^{-j}_m{]}={[}U_n,Q^a_r{]}={[}U_n,A^{\pm i}_m{]}
\label{LN4}
\eea
$\Phi$ is any of the generators $G,A,U$ or $Q$. 

\section{Non-reductive $N=4$ superconformal algebra}

Let us introduce the linearly combined generators
\bea 
 G^{\pm\al}=\hf(G^1\pm iG^2)\ ,&&\ \ \ \ \ G^{\pm\beta}=\hf(G^3\pm iG^4)
  \nn 
 Q^{\pm\al}=\hf(Q^1\pm iQ^2)\ ,&&\ \ \ \ \ Q^{\pm\beta}=\hf(Q^3\pm iQ^4)
\label{al}
\eea
and
\ben
 E^\pm=A^{\pm 2}-iA^{\pm 1}\ ,\ \ \ H^\pm=2iA^{\pm3}\ ,\ \ \ F^\pm=
   -A^{\pm2}-iA^{\pm1}
\een
We immediately rescale and rename some of the generators:
\bea
 &&\phi^{-\al}=2\g Q^{-\al}\ ,\ \ \ \ \phi^{-\beta}=2\g Q^{-\beta}\nn
 &&\phi^\al=2Q^\al\ ,\ \ \ \ \ \phi^\beta=2Q^\beta\nn
 &&J=E^+\ ,\ \ \ \ V=\g H^+\ ,\ \ \ \ R=\g F^+\nn
 &&E=E^-\ ,\ \ \ \ H=H^-\ ,\ \ \ \ F=F^- 
\label{scale1}
\eea
Note that $E^+$ is not scaled.
We also replace $U$ with the spin-1 generator $W$:
\ben
 W=U-\hf H^+
\label{W}
\een 
whereby the generator briefly known as $H^+$ appears scaled as well as 
unscaled. The remaining five generators $L$ and $G$
are left unscaled. The conformal weights
are ``inherited'' by the new generators (\ref{al}) to (\ref{W}):
$\D(\Phi)\in{\{}3/2,1,\dots,1,1/2{\}}$
for $\Phi\in{\{}G^{\pm\al,\beta},E,H,F,J,V,R,W,\phi^{\pm\al,\beta}{\}}$.

In terms of this equivalent set of generators, the non-vanishing 
(anti-)commutators of the large $N=4$ SCA are
\bea
 &&{[}L_n,L_m{]}=(n-m)L_{n+m}+\frac{c}{12}(n^3-n)\delta_{n+m,0}\nn
 &&{[}L_n,\Phi_r{]}=((\Delta(\Phi)-1)n-r)\Phi_{n+r}\nn
 &&\{G^\al_r,G^{-\al}_s\}=L_{r+s}+(r-s)\left(\hf V_{r+s}+\hf(1-\g)H_{r+s}
   \right)+\frac{c}{6}(r^2-1/4)\delta_{r+s,0}\nn
 &&\{G^\beta_r,G^{-\beta}_s\}=L_{r+s}+(r-s)\left(\hf V_{r+s}-\hf(1-\g)H_{r+s}
    \right)+\frac{c}{6}(r^2-1/4)\delta_{r+s,0}\nn
 &&\{G^\al_r,G^{\beta}_s\}=(r-s)\g J_{r+s}\ ,\ \ \ \
  \{G^{-\al}_r,G^{-\beta}_s\}=-(r-s)R_{r+s}\nn
 &&\{G^\al_r,G^{-\beta}_s\}=(r-s)(1-\g)E_{r+s}\ ,\ \ \ \
  \{G^\beta_r,G^{-\al}_s\}=(r-s)(1-\g)F_{r+s}\nn
 &&{[}E_n,G_r^{-\al}{]}=-G^{-\beta}_{n+r}-n\phi^{-\beta}_{n+r}\ ,\ \ \ \
  {[}E_n,G_r^{\beta}{]}=G^{\al}_{n+r}+\g n\phi^{\al}_{n+r}\nn
 &&{[}H_n,G_r^{\al}{]}=G^\al_{n+r}+\g n\phi^\al_{n+r}\ ,\ \ \ \
  {[}H_n,G_r^{-\al}{]}=-G^{-\al}_{n+r}-n\phi^{-\al}_{n+r}\nn
 &&{[}H_n,G_r^{\beta}{]}=-G^\beta_{r+n}-\g n\phi^\beta_{n+r}\ ,\ \ \ \
  {[}H_n,G_r^{-\beta}{]}=G^{-\beta}_{n+r}+n\phi^{-\beta}_{n+r}\nn
 &&{[}F_n,G_r^{\al}{]}=G^{\beta}_{n+r}+\g n\phi^{\beta}_{n+r}\ ,\ \ \ \
  {[}F_n,G_r^{-\beta}{]}=-G^{-\al}_{n+r}-n\phi^{-\al}_{n+r}\nn
 &&{[}H_n,E_m{]}=2E_{n+m}\ ,\ \ \ \
  {[}H_n,F_m{]}=-2F_{n+m}\nn
 &&{[}H_n,H_m{]}=\frac{c}{3(1-\g)}n\delta_{n+m,0}\ ,\ \ \ \
  {[}E_n,F_m{]}=H_{n+m}+\frac{c}{6(1-\g)}n\delta_{n+m,0}\nn
 &&{[}E_n,\phi_r^{-\al}{]}=-\phi^{-\beta}_{n+r}\ ,\ \ \ \
  {[}E_n,\phi_r^{\beta}{]}=\phi^{\al}_{n+r}\nn
 &&{[}H_n,\phi_r^{\al}{]}=\phi^\al_{n+r}\ ,\ \ \ \
  {[}H_n,\phi_r^{-\al}{]}=-\phi^{-\al}_{n+r}\nn
 &&{[}H_n,\phi_r^{\beta}{]}=-\phi^\beta_{r+n}\ ,\ \ \ \
  {[}H_n,\phi_r^{-\beta}{]}=\phi^{-\beta}_{n+r}\nn
 &&{[}F_n,\phi_r^{\al}{]}=\phi^{\beta}_{n+r}\ ,\ \ \ \
  {[}F_n,\phi_r^{-\beta}{]}=-\phi^{-\al}_{n+r}\nn
 &&{[}J_n,G_r^{-\al}{]}=-G^\beta_{n+r}+(1-\g)n\phi^\beta_{n+r}\ ,\ \ \ \
  {[}J_n,G_r^{-\beta}{]}=G^\al_{n+r}-(1-\g)n\phi^\al_{n+r}\nn
 &&{[}V_n,G_r^{\al}{]}=\g G^\al_{n+r}-\g(1-\g)n\phi^\al_{n+r}\ ,\ \ \ \
  {[}V_n,G_r^{-\al}{]}=-\g G^{-\al}_{n+r}+(1-\g)n\phi^{-\al}_{n+r}\nn
 &&{[}V_n,G_r^{\beta}{]}=\g G^\beta_{r+n}-\g(1-\g)n\phi^\beta_{n+r}\ ,\ \ \ \
  {[}V_n,G_r^{-\beta}{]}=-\g G^{-\beta}_{n+r}+(1-\g)n\phi^{-\beta}_{n+r}\nn
 &&{[}R_n,G_r^{\al}{]}=\g G^{-\beta}_{n+r}-(1-\g)n\phi^{-\beta}_{n+r}\ ,
   \ \ \ \
  {[}R_n,G_r^{\beta}{]}=-\g G^{-\al}_{n+r}+(1-\g)n\phi^{-\al}_{n+r}\nn
 &&{[}V_n,J_m{]}=2\g J_{n+m}\ ,\ \ \ \
  {[}V_n,R_m{]}=-2\g R_{n+m}\nn
 &&{[}V_n,V_m{]}=\frac{c}{3}\g n\delta_{n+m,0}\ ,\ \ \ \
  {[}J_n,R_m{]}=V_{n+m}+\frac{c}{6}n\delta_{n+m,0}\nn
 &&{[}J_n,\phi_r^{-\al}{]}=-\g\phi^{\beta}_{n+r}\ ,\ \ \ \
  {[}J_n,\phi_r^{-\beta}{]}=\g\phi^{\al}_{n+r}\nn
 &&{[}V_n,\phi_r^{\al}{]}=\g\phi^\al_{n+r}\ ,\ \ \ \
  {[}V_n,\phi_r^{-\al}{]}=-\g \phi^{-\al}_{n+r}\nn
 &&{[}V_n,\phi_r^{\beta}{]}=\g\phi^\beta_{r+n}\ ,\ \ \ \
  {[}V_n,\phi_r^{-\beta}{]}=-\g\phi^{-\beta}_{n+r}\nn
 &&{[}R_n,\phi_r^{\al}{]}=\phi^{-\beta}_{n+r}\ ,\ \ \ \
  {[}R_n,\phi_r^{\beta}{]}=-\phi^{-\al}_{n+r}\nn
 &&\{\phi^\al_r,G^{-\al}_s\}=W_{r+s}+\frac{1}{2}H_{r+s}\ ,\ \ \ \
  \{\phi^\al_r,G^\beta_s\}=-J_{r+s}\nn
 &&\{\phi^\al_r,G^{-\beta}_s\}=E_{r+s}\ ,\ \ \ \
  \{\phi^\beta_r,G^\al_s\}=J_{r+s}\nn
 &&\{\phi^\beta_r,G^{-\al}_s\}=F_{r+s}\ ,\ \ \ \
  \{\phi^\beta_r,G^{-\beta}_s\}=W_{r+s}-\frac{1}{2}H_{r+s}\nn
 &&\{\phi^{-\al}_r,G^\al_s\}=V_{r+s}+\g W_{r+s}-\hf\g H_{r+s}\ ,\ \ \ \
  \{\phi^{-\al}_r,G^\beta_s\}=-\g F_{r+s}\nn
 &&\{\phi^{-\al}_r,G^{-\beta}_s\}=R_{r+s}\ ,\ \ \ \
  \{\phi^{-\beta}_r,G^\al_s\}=-\g E_{r+s}\nn
 &&\{\phi^{-\beta}_r,G^{-\al}_s\}=-R_{r+s}\ ,\ \ \ \
  \{\phi^{-\beta}_r,G^{\beta}_s\}=V_{r+s}+\g W_{r+s}+\hf\g H_{r+s}\nn
 &&\{\phi^\al_r,\phi^{-\al}_s\}=-\frac{c}{6(1-\g)}\delta_{r+s,0}\ ,\ \ \ \
  \{\phi^\beta_r,\phi^{-\beta}_s\}=-\frac{c}{6(1-\g)}\delta_{r+s,0}\nn
 &&{[}W_n,G^\al_r{]}=-\hf G^{\al}_{n+r}+(1-\g/2)n\phi^{\al}_{n+r}\ , \ \ \ \
  {[}W_n,G^{-\al}_r{]}=\hf G^{-\al}_{n+r}+\hf n\phi^{-\al}_{n+r}\nn
 &&{[}W_n,G^\beta_r{]}=-\hf G^{\beta}_{n+r}+(1-\g/2)n\phi^{\beta}_{n+r}\ , 
    \ \ \ \
  {[}W_n,G^{-\beta}_r{]}=\hf G^{-\beta}_{n+r}+\hf n\phi^{-\beta}_{n+r}\nn
 &&{[}W_n,J_m{]}=-J_{n+m}\ ,\ \ \ \
  {[}W_n,V_m{]}=-\frac{c}{6}n\delta_{n+m,0}\ ,\ \ \ \
  {[}W_n,R_m{]}=R_{n+m}\nn
 &&{[}W_n,\phi^\al_r{]}=-\hf\phi^{\al}_{n+r}\ ,\ \ \ \
  {[}W_n,\phi^{-\al}_r{]}=\hf\phi^{-\al}_{n+r}\nn
 &&{[}W_n,\phi^\beta_r{]}=-\hf\phi^{\beta}_{n+r}\ ,\ \ \ \
  {[}W_n,\phi^{-\beta}_r{]}=\hf\phi^{-\beta}_{n+r}\nn
 &&{[}W_n,W_m{]}=-\frac{c}{12(1-\g)}n\delta_{n+m,0}
\label{gSCA}
\eea
As usual, we let $\Phi$ denote any of the 15 generators different from $L$.

\subsection{In\"on\"u-Wigner contraction}

The IW contraction of our interest corresponds to considering the
limit $\g\rightarrow0$ of the large $N=4$ SCA in the form (\ref{gSCA}).
Even though the limit appears singular from the point of view of the
redefinitions (\ref{scale1}), the algebra (\ref{gSCA}) does not
display any divergencies. It merely reduces to
\bea
 &&{[}L_n,L_m{]}=(n-m)L_{n+m}+\frac{c}{12}(n^3-n)\delta_{n+m,0}\nn
 &&{[}L_n,\Phi_r{]}=((\Delta(\Phi)-1)n-r)\Phi_{n+r}\nn
 &&\{G^\al_r,G^{-\al}_s\}=L_{r+s}+\hf(r-s)\left(V_{r+s}+H_{r+s}
   \right)+\frac{c}{6}(r^2-1/4)\delta_{r+s,0}\nn
 &&\{G^\beta_r,G^{-\beta}_s\}=L_{r+s}+\hf(r-s)\left(V_{r+s}-H_{r+s}
    \right)+\frac{c}{6}(r^2-1/4)\delta_{r+s,0}\nn
 &&\{G^\al_r,G^{-\beta}_s\}=(r-s)E_{r+s}\ ,\ \ \ \
  \{G^\beta_r,G^{-\al}_s\}=(r-s)F_{r+s}\ ,\ \ \ \
  \{G^{-\al}_r,G^{-\beta}_s\}=-(r-s)R_{r+s}\nn
 &&{[}E_n,G_r^{-\al}{]}=-G^{-\beta}_{n+r}-n\phi^{-\beta}_{n+r}\ ,\ \ \ \
  {[}E_n,G_r^{\beta}{]}=G^{\al}_{n+r}\nn
 &&{[}H_n,G_r^{\al}{]}=G^\al_{n+r}\ ,\ \ \ \
  {[}H_n,G_r^{-\al}{]}=-G^{-\al}_{n+r}-n\phi^{-\al}_{n+r}\nn
 &&{[}H_n,G_r^{\beta}{]}=-G^\beta_{r+n}\ ,\ \ \ \
  {[}H_n,G_r^{-\beta}{]}=G^{-\beta}_{n+r}+n\phi^{-\beta}_{n+r}\nn
 &&{[}F_n,G_r^{\al}{]}=G^{\beta}_{n+r}\ ,\ \ \ \
  {[}F_n,G_r^{-\beta}{]}=-G^{-\al}_{n+r}-n\phi^{-\al}_{n+r}\nn
 &&{[}H_n,E_m{]}=2E_{n+m}\ ,\ \ \ \
  {[}H_n,F_m{]}=-2F_{n+m}\nn
 &&{[}H_n,H_m{]}=\frac{c}{3}n\delta_{n+m,0}\ ,\ \ \ \
  {[}E_n,F_m{]}=H_{n+m}+\frac{c}{6}n\delta_{n+m,0}\nn
 &&{[}E_n,\phi_r^{-\al}{]}=-\phi^{-\beta}_{n+r}\ ,\ \ \ \
  {[}H_n,\phi_r^{-\al}{]}=-\phi^{-\al}_{n+r}\nn
 &&{[}H_n,\phi_r^{-\beta}{]}=\phi^{-\beta}_{n+r}\ ,\ \ \ \
  {[}F_n,\phi_r^{-\beta}{]}=-\phi^{-\al}_{n+r}\nn
 &&{[}V_n,G_r^{-\al}{]}=n\phi^{-\al}_{n+r}\ ,\ \ \ \
  {[}V_n,G_r^{-\beta}{]}=n\phi^{-\beta}_{n+r}\nn
 &&{[}R_n,G_r^{\al}{]}=-n\phi^{-\beta}_{n+r}\ ,\ \ \ \
  {[}R_n,G_r^{\beta}{]}=n\phi^{-\al}_{n+r}\nn
 &&\{\phi^{-\al}_r,G^\al_s\}=V_{r+s}\ ,\ \ \ \
  \{\phi^{-\al}_r,G^{-\beta}_s\}=R_{r+s}\nn
 &&\{\phi^{-\beta}_r,G^{-\al}_s\}=-R_{r+s}\ ,\ \ \ \
  \{\phi^{-\beta}_r,G^{\beta}_s\}=V_{r+s}
\label{red}
\eea
and
\bea
 &&{[}E_n,\phi_r^{\beta}{]}=\phi^{\al}_{n+r}\ ,\ \ \ \
  {[}H_n,\phi_r^{\al}{]}=\phi^\al_{n+r}\nn
 &&{[}H_n,\phi_r^{\beta}{]}=-\phi^\beta_{r+n}\ ,\ \ \ \
  {[}F_n,\phi_r^{\al}{]}=\phi^{\beta}_{n+r}\nn
 &&{[}J_n,G_r^{-\al}{]}=-G^\beta_{n+r}+n\phi^\beta_{n+r}\ ,\ \ \ \
  {[}J_n,G_r^{-\beta}{]}=G^\al_{n+r}-n\phi^\al_{n+r}\nn
 &&{[}J_n,R_m{]}=V_{n+m}+\frac{c}{6}n\delta_{n+m,0}\nn
 &&{[}R_n,\phi_r^{\al}{]}=\phi^{-\beta}_{n+r}\ ,\ \ \ \
  {[}R_n,\phi_r^{\beta}{]}=-\phi^{-\al}_{n+r}\nn
 &&\{\phi^\al_r,G^{-\al}_s\}=W_{r+s}+\frac{1}{2}H_{r+s}\ ,\ \ \ \
  \{\phi^\al_r,G^\beta_s\}=-J_{r+s}\ ,\ \ \ \
  \{\phi^\al_r,G^{-\beta}_s\}=E_{r+s}\nn
 &&\{\phi^\beta_r,G^\al_s\}=J_{r+s}\ ,\ \ \ \
  \{\phi^\beta_r,G^{-\al}_s\}=F_{r+s}\ ,\ \ \ \
  \{\phi^\beta_r,G^{-\beta}_s\}=W_{r+s}-\frac{1}{2}H_{r+s}\nn
 &&\{\phi^\al_r,\phi^{-\al}_s\}=-\frac{c}{6}\delta_{r+s,0}\ ,\ \ \ \
  \{\phi^\beta_r,\phi^{-\beta}_s\}=-\frac{c}{6}\delta_{r+s,0}\nn
 &&{[}W_n,G^\al_r{]}=-\hf G^{\al}_{n+r}+n\phi^{\al}_{n+r}\ ,\ \ \ \
  {[}W_n,G^{-\al}_r{]}=\hf G^{-\al}_{n+r}+\hf n\phi^{-\al}_{n+r}\nn
 &&{[}W_n,G^\beta_r{]}=-\hf G^{\beta}_{n+r}+n\phi^{\beta}_{n+r}\ ,\ \ \ \
  {[}W_n,G^{-\beta}_r{]}=\hf G^{-\beta}_{n+r}+\hf n\phi^{-\beta}_{n+r}\nn
 &&{[}W_n,J_m{]}=-J_{n+m}\ ,\ \ \ \
  {[}W_n,V_m{]}=-\frac{c}{6}n\delta_{n+m,0}\ ,\ \ \ \
  {[}W_n,R_m{]}=R_{n+m}\nn
 &&{[}W_n,\phi^\al_r{]}=-\hf\phi^{\al}_{n+r}\ ,\ \ \ \
  {[}W_n,\phi^{-\al}_r{]}=\hf\phi^{-\al}_{n+r}\nn
 &&{[}W_n,\phi^\beta_r{]}=-\hf\phi^{\beta}_{n+r}\ ,\ \ \ \
  {[}W_n,\phi^{-\beta}_r{]}=\hf\phi^{-\beta}_{n+r}\nn
 &&{[}W_n,W_m{]}=-\frac{c}{12}n\delta_{n+m,0}
\label{red2}
\eea
This algebra is singly extended, parameterised by the central charge $c=6k^-$.
For reasons which will become clear shortly, 
we have written the many non-vanishing
(anti-)commutators in two separate families: (\ref{red}) and (\ref{red2}).
Half-integer moding of the fermionic generators $G$ and $\phi$
corresponds to a
Neveu-Schwarz sector, while integer moding corresponds to a Ramond sector.

The Jacobi identities (\ref{Jacobi}) are in general not ensured
after an IW contraction. However, we have carried out explicitly the tedious
job of verifying (\ref{Jacobi}) for the 16 generators of
(\ref{red}) and (\ref{red2}). Hence, this new SCA is well-defined.
It is asymmetric in the way the supercurrents are treated,
since $\{G^{-\al}_r,G^{-\beta}_s\}=-(r-s)R_{r+s}$ while
$\{G^\al_r,G^\beta_s\}=0$. This interesting feature was also
present in the SCA of \cite{Ras2}.

We observe that (\ref{red}) is a subalgebra of the full SCA.
When compared to the asymmetric SCA of \cite{Ras2}, this subalgebra
is equivalent to the asymmetric one,
provided the spin-1 generator $R$ is considered the
derivative of the spin-0 generator $S$ of \cite{Ras2}:
\ben 
 R_n=nS_n
\een
This means that the asymmetric SCA is slightly bigger than (\ref{red})
since certain commutators involving $S_0$ are non-vanishing \cite{Ras2}.

In the extension governed by (\ref{red2}), we can not consider
the spin-1 field $R$ straightforwardly as the derivative of a spin-0
field $S$. This follows from the commutator $[J_n,R_m]$, for example,
where the right hand side involves the spin-1 field $V$.

Equivalent, non-reductive $N=4$ SCAs may be constructed by similar
IW contractions. First, scaling $E^+$ and $Q^+$ instead of $F^+$ and
$Q^-$ (\ref{scale1}) will lead to an isomorphic SCA with the roles
of $G^+$ and $G^-$ interchanged. Similarly, one could consider the limit
$\g\rightarrow1$ in which case one would have to scale $H^-$ and
either $F^-$ or $E^-$ (and leave $H^+$, $E^+$ and $F^+$ unscaled). 
In this limit the central extension becomes $c=6k^+$.

\section{Affine extensions of Lie algebras}

Let us conclude by classifying the possible affine extensions of the
four-dimensional and non-reductive Lie algebra $g$. It is generated
by the zero-modes $\{J_0,V_0,R_0,W_0\}$ of the particular
affine Lie algebra appearing in our construction above:
\bea
 {[}J_n,R_m{]}&=&V_{n+m}+\frac{c}{6}n\delta_{n+m,0}\nn
 {[}W_n,J_m{]}&=&-J_{n+m}\nn
 {[}W_n,V_m{]}&=&-\frac{c}{6}n\delta_{n+m,0}\nn
 {[}W_n,R_m{]}&=&R_{n+m}\nn
 {[}W_n,W_m{]}&=&-\frac{c}{12}n\delta_{n+m,0}\nn
 0&=&{[}J_n,J_m{]}={[}V_n,J_m{]}={[}V_n,V_m{]}={[}V_n,R_m{]}={[}R_n,R_m{]}
\label{four}
\eea
Abbreviating the zero-modes by $\{J,V,R,W\}$, the non-vanishing commutators
of the algebra $g$ are
\ben
 {[}J,R{]}=V\ ,\ \ \ \ {[}W,J{]}=-J\ ,\ \ \ \ {[}W,R{]}=R
\label{g}
\een
This algebra is recognized as the semi-direct sum
\ben
 g=u(1)\ \bar\oplus\ n_3
\een
where the Lie algebra $n_3$ of strictly upper-triangular $3\times3$-matrices
is the biggest non-trivial ideal of $g$. The $u(1)$ is generated by
$\{W\}$, while $n_3$ is generated by $\{J,V,R\}$. $g$ is seen to be
solvable.

An affine extension of the generic Lie algebra
\ben
 [j_a,j_b]={f_{ab}}^c j_c
\een
is governed by the central extension (or level) $k$, and a symmetric,
bilinear and invariant two-form $\kappa$:
\bea
 \kappa_{ab}&=&\kappa(j_a,j_b)\nn
 \kappa({[}j_a,j_b{]},j_c)&=&\kappa(j_a,{[}j_b,j_c{]})
\label{kappa}
\eea
The resulting affine Lie algebra reads
\ben
 [j_{a,n},j_{b,m}]={f_{ab}}^c j_{c,n+m}+nk\kappa_{ab}\delta_{n+m,0}
\label{affine}
\een
The invariance of $\kappa$, in particular, is required by the Jacobi 
identities of (\ref{affine}). Note that $k$ may be absorbed in a rescaling
of $\kappa$. However, it is convenient to factorize the extension into
the Lie algebra dependent object $\kappa$, and the purely affine entity $k$.
We see that a classification of possible affine extensions of a given
Lie algebra amounts to classifying the $\kappa$-forms of the Lie algebra.

The canonical choice of $\kappa$-form (\ref{kappa}) is the ordinary
Cartan-Killing form
\ben
 \kappa_{ab}={f_{ac}}^d{f_{bd}}^c
\label{CK}
\een
For simple Lie algebras, it is unique up to an overall scaling (the 
normalization chosen here is unconventional but irrelevant to our purpose).
For non-reductive Lie algebras, on the other hand, there will in general
exist several inequivalent $\kappa$-forms. Here we shall classify them in 
the case of $g$ (\ref{g}).

The approach is straightforward and can be applied to any finite-dimensional
Lie algebra. The defining properties of the $\kappa$-form implies the
anti-symmetry
\ben
     f_{cab}=-f_{cba}\ ,\ \ \ \ {\rm where} \ \ f_{cab}={f_{ca}}^d\kappa_{db}
\label{fk}
\een
It should be noted that since $\kappa$ may be degenerate, 
it in general can not be used as a metric whose inverse can raise indices. 
Now, considering $f_{cab}$ as a matrix element of the $d\times d$-matrix
$f_c$ ($d$ is the dimension of the Lie algebra), the anti-symmetry
(\ref{fk}) imposes constraints on $\kappa$. Along with the symmetry 
of $\kappa$, $\kappa_{ab}=\kappa_{ba}$,
those are the only constraints to impose. Thus,
the classification is achieved by an analysis of the $d$ matrices $f_c$.
In the case of $g$ (\ref{g}) we find the general $\kappa$-form
\ben
 \kappa=\la\left(\begin{array}{llll}0&0&0&0\\ 0&0&0&0\\ 0&0&0&0\\ 0&0&0&1
  \end{array}\right)
   +\mu\left(\begin{array}{llll}0&0&1&0\\ 0&0&0&\bar1\\ 1&0&0&0\\ 0&\bar1&0&0
  \end{array}\right)
\label{lmn}
\een
Here $\bar1\equiv-1$.
The matrix elements are labelled according to the order $\{J,V,R,W\}$,
i.e., $\kappa_{24}=\kappa(V,W)$ etc.
We see that the ordinary Cartan-Killing form corresponds to 
$(\la,\mu)=(2,0)$, while the affine extension (\ref{four})
corresponds to $(\la,\mu)=(-1,2)$. In the latter case, the level has been
normalized to $k=c/12$.

As a final comment, we observe that in terms of the linear combinations
\ben
 U^+_n=V_n-2W_n\ ,\ \ \ \ U^-_n=-2W_n
\een
(\ref{four}) becomes
\bea
 {[}J_n,R_m{]}&=&U^+_{n+m}-U^-_{n+m}+\frac{c}{6}n\delta_{n+m,0}\nn
 {[}U^\pm_n,J_m{]}&=&2J_{n+m}\nn
 {[}U^\pm_n,R_m{]}&=&-2R_{n+m}\nn
 {[}U^\pm_n,U^\pm_m{]}&=&\pm\frac{c}{3}n\delta_{n+m,0}\nn
 0&=&{[}U^+_n,U^-_m{]}={[}J_n,J_m{]}={[}R_n,R_m{]}
\label{four2}
\eea
Thus, the affine Lie algebra (\ref{four}) is seen to contain
a level $k=c/6$ affine $su(2)$ Lie algebra generated
by $\{J,U^+,R\}$, and a level $k=-c/6$ affine $su(2)$ Lie algebra
generated by $\{-J,U^-,R\}$ (or equivalently by $\{J,U^-,-R\}$).
We have used the term {\em contain} deliberately to emphasize that they
are obviously {\em not} subalgebras.

\vskip.5cm
\noindent{\em Acknowledgements}
\vskip.1cm
\noindent The author thanks Mark Walton for stimulating discussions and
comments on the manuscript,
and Bruce Campbell for encouragement. The author is also grateful to CRM, 
Montreal, for its generous hospitality, and he is
supported in part by a PIMS Postdoctoral Fellowship and by NSERC.
In addendum, the author thanks the referee for his insight pointing
out several misprints, and Antoine Van Proeyen for a clarifying comment.

\end{document}